\documentclass[%
superscriptaddress,
twocolumn,
showpacs,preprintnumbers,
amsmath,amssymb,
aps,
]{revtex4-1}

\usepackage{graphicx}
\usepackage{dcolumn}
\usepackage{bm}
\usepackage[mathlines]{lineno}
\usepackage{hyperref}
\hypersetup{
    colorlinks=true,
    linkcolor=blue,
    filecolor=blue,
    urlcolor=blue,
    citecolor=blue,
}

\begin{document}

\title{Unique electronic state in ferromagnetic semiconductor FeCl$_{2}$ monolayer}

\author{Di Lu}
\affiliation{Laboratory for Computational Physical Sciences (MOE),
	State Key Laboratory of Surface Physics, and Department of Physics,
	Fudan University, Shanghai 200433, China}
\affiliation{Shanghai Qi Zhi Institute, Shanghai 200232, China}

\author{Lu Liu}
\affiliation{Laboratory for Computational Physical Sciences (MOE),
	State Key Laboratory of Surface Physics, and Department of Physics,
	Fudan University, Shanghai 200433, China}
\affiliation{Shanghai Qi Zhi Institute, Shanghai 200232, China}

\author{Yaozhenghang Ma}
\affiliation{Laboratory for Computational Physical Sciences (MOE),
	State Key Laboratory of Surface Physics, and Department of Physics,
	Fudan University, Shanghai 200433, China}
\affiliation{Shanghai Qi Zhi Institute, Shanghai 200232, China}

\author{Ke Yang}
\affiliation{College of Science, University of Shanghai for Science and Technology,
       Shanghai 200093, China}
       \affiliation{Laboratory for Computational Physical Sciences (MOE),
	State Key Laboratory of Surface Physics, and Department of Physics,
	Fudan University, Shanghai 200433, China}

\author{Hua Wu}
\affiliation{Laboratory for Computational Physical Sciences (MOE),
	State Key Laboratory of Surface Physics, and Department of Physics,
	Fudan University, Shanghai 200433, China}
\affiliation{Shanghai Qi Zhi Institute, Shanghai 200232, China}
\affiliation{Collaborative Innovation Center of Advanced Microstructures,
	Nanjing 210093, China}

\begin{abstract}
Two-dimensional (2D) van der Waals (vdW) magnetic materials could be an ideal platform for ultracompact spintronic applications. Among them, FeCl$_{2}$ monolayer in the triangular lattice is subject to a strong debate. Thus, we critically examine its spin-orbital state, electronic structure, and magnetic properties, using a set of delicate first-principles calculations, crystal field level analyses, and Monte Carlo simulations. Our work reveals that FeCl$_{2}$ monolayer is a ferromagnetic (FM) semiconductor in which the electron correlation of the narrow Fe $3d$ bands determines the band gap of about 1.2 eV. Note that only when the spin-orbit coupling (SOC) is properly handled, the unique $d$$^{5\uparrow}$$l$$^\downarrow_{z+}$ electronic ground state is achieved. Then, both the orbital and spin contributions (0.59 $\mu_{\rm B}$ plus 3.56 $\mu_{\rm B}$) to the total magnetic moment well account for, for the first time, the experimental perpendicular moment of 4.3 $\mu_{\rm B}$/Fe. Moreover, we find that a compressive strain further stabilizes the $d$$^{5\uparrow}$$l$$^\downarrow_{z+}$ ground state, and that the enhanced magnetic anisotropy and exchange coupling would boost the Curie temperature ($T_{\rm C}$) from 25 K for the pristine FeCl$_{2}$ monolayer to 69-102 K under 3$\%$-5$\%$ compressive strain. Therefore, FeCl$_{2}$ monolayer is indeed an appealing 2D FM semiconductor.
\end{abstract}

\maketitle

\section{INTRODUCTION}
Reduced dimensionality and interlayer couplings of bulk vdW materials trigger intriguing electronic, optical and other quantum properties~\cite{novoselov2004,novoselov2005,zhang2005,cao2018,sun2019}. According to the Mermin-Wagner theorem~\cite{mermin1966}, long-range magnetic order at finite temperature is prohibited in 2D isotropic Heisenberg spin-rotational-invariant systems. However, magnetic anisotropies can break spin-rotation symmetry, thus bringing about possible long-range magnetism. Recently, 2D vdW magnets with tunable magnetic anisotropy have attracted a large volume of attention as 2D FM was observed in atomically thin CrI$_3$~\cite{huang2017} and Cr$_2$Ge$_2$Te$_6$~\cite{gong2017}. These 2D vdW magnets exhibit appealing properties, such as the magnetic anisotropy~\cite{kim2019,yang2020,liu2020}, the exotic quantum spin liquid states~\cite{xu2020} and aniferromagnetic (AF) topological insulators~\cite{gong2019,deng2020,otrokov2019}. The abundant properties open a new avenue to spintronic applications, such as spin valves~\cite{cardoso2018}, spin filters~\cite{klein2018,song2018}, and data storage~\cite{soumyanarayanan2016}. Moreover, owing to the layered structure, one could be able to control the 2D magnetic properties by strain~\cite{yang2020,liu2020}, doping~\cite{deng2018,huang2018}, heterostructure~\cite{gibertini2019,liushan2020} or applying magnetic/electric field~\cite{huang018,jiang2018}. For example, the $T_{\rm C}$ of monolayer Fe$_3$GeTe$_2$ can exceed room temperature via ionic gating.
\begin{figure}[t]
\centering
	\includegraphics[width=9cm]{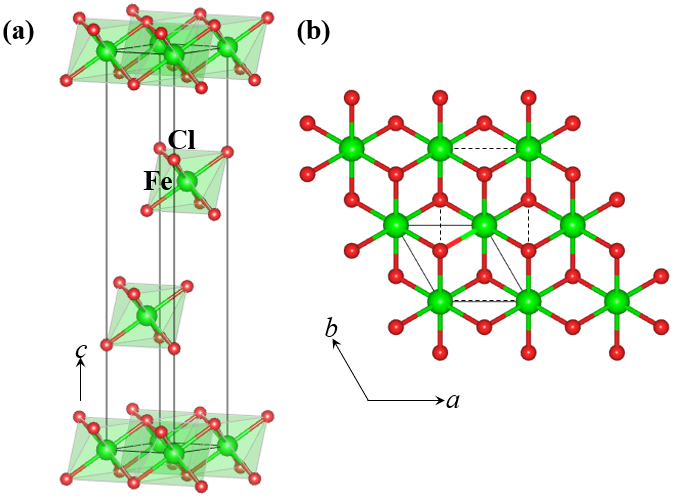}
	\caption {(a) The crystal structure of FeCl$_{2}$ bulk and (b) FeCl$_{2}$ monolayer: the solid diamond represents the unit cell, and the dashed rectangular $1\times\sqrt{3}$ supercell models the stripe AF state.}
	\label{fig1}
\end{figure}

The binary transition metal dihalides MX$_2$ (M = transition metal; X = halogen: Cl, Br, I) are vdW layered materials containing a triangular lattice of M$^{2+}$ cations~\cite{mcguire2017}. The bulk magnetism was analyzed several decades ago~\cite{wilkinson1959,lines1963,birgeneau1972,jacobs1967}. Very recently, FeCl$_2$ monolayer has been synthesized by molecular-beam epitaxy~\cite{zhou2020,cai2020}, and it is a new 2D vdW magnet with a semiconducting gap of about 1.2 eV~\cite{cai2020}. However, several theoretical studies suggest that it is a high-spin ($S$ = 2) half-metallic (HM) ferromagnet with a conducting down-spin $t_{2g}$ channel in the local octahedral coordination~\cite{torun2015,ashton2017,feng2018,kulish2017,ghosh2021}. Another two studies present the FM insulating solution~\cite{botana2019,yao2021}, but the ground state was not yet firmly determined through total energy calculations. Moreover, the orbital singlet solution $d$$^{5\uparrow}$$a$$^\downarrow_{1g}$ was obtained when the reasonable $U$ and SOC parameters were used~\cite{yao2021}. Then, only the spin moment is expected, and it should be smaller than 4 $\mu_{\rm B}$ per Fe$^{2+}$ ($S$ = 2) reduced by the Fe-Cl covalence; and a small orbital moment, if any, would be in the plane (see more below, e.g. Table 1). Note that FeCl$_2$ bulk was reported to have interlayer AF but intralayer FM couplings, with the N\'eel temperature $T_{\rm N}$ = 24 K and a perpendicular magnetic moment of 4.3 $\mu_{\rm B}$ exceeding the spin-only value~\cite{wilkinson1959,birgeneau1972,jacobs1967}. So far, the intralayer FM coupling of most concern for FeCl$_2$ monolayer has been discussed in the FM HM solution in the previous studies~\cite{ashton2017,kulish2017,ghosh2021}, in which the $T_{\rm C}$ was too much overestimated by the fictitious itinerant magnetism, instead of the true superexchange in this FM semiconducting monolayer. Therefore, the FM semiconducting behavior of FeCl$_2$ monolayer and the perpendicular magnetic moment remain largely unresolved.
\begin{figure}[t]
\centering
	\includegraphics[width=8cm]{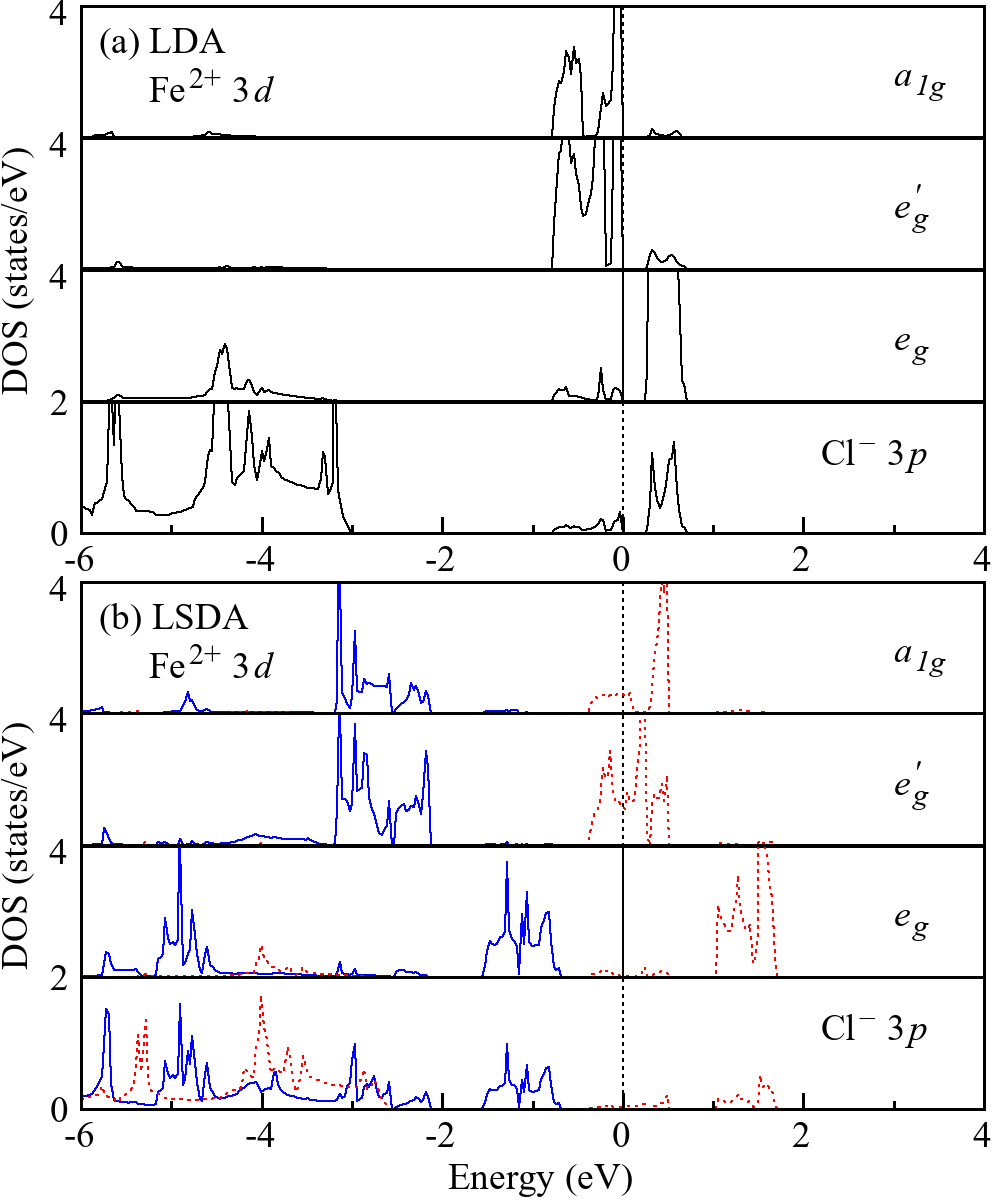}
	\caption {The DOS results of (a) LDA and (b) LSDA for FeCl$_{2}$ monolayer. The blue (red) curves refer to the majority (minority) spin. The Fermi level is set at the zero energy.}
    \label{fig2}
\end{figure}

In this work, we critically examine the electronic and magnetic structures of FeCl$_2$ monolayer using a set of delicate first-principles calculations, crystal field level analyses, and Monte Carlo simulations. Our calculations find a moderate $t_{2g}$-$e_g$ crystal field splitting for the Fe $3d$ states, and thus a high-spin ($S$ = 2) Fe$^{2+}$ state is achieved by the Hund exchange. Moreover, under the global trigonal crystal field, Fe $3d$ $t_{2g}$ triplet splits and the $e'_g$ doublet is lower than $a_{1g}$ singlet by 30 meV. Owing to the strong correlation effect of the narrow Fe $3d$ bands in the triangular lattice, the band gap is determined by the Hubbard $U$. But note that here the orbital states must be delicately handled, and only when the SOC is properly included, the correct electronic ground state $d$$^{5\uparrow}$$l$$^\downarrow_{z+}$ would be achieved. Then, an out-of-plane orbital moment, strong perpendicular magnetic anisotropy of single ion type, and intralayer FM coupling, all well account for the FM semiconducting behavior of FeCl$_2$ monolayer, with $T_{\rm C}$ = 25 K being estimated by our Monte Carlo simulations. Furthermore, we predict that the $d$$^{5\uparrow}$$l$$^\downarrow_{z+}$ ground state can be further stabilized by a compressive strain, and --3\% (--5\%) strain would raise the $T_{\rm C}$ to 69 K (102 K). Therefore, FeCl$_2$ monolayer is indeed an appealing 2D FM semiconductor.

\section{COMPUTATIONAL DETAILS}
Density functional theory (DFT) calculations were carried out using the full-potential augmented plane wave plus local orbital code (Wien2k)~\cite{blaha2001}. The optimized lattice parameter of FeCl$_2$ monolayer is $a$ = $b$ = 6.708 \AA, which is close to (within 1.5\%) the experimental value of 6.809 \AA~for the bulk~\cite{wilkinson1959}. A vacuum slab of 15 \AA~was set along the $c$-axis. The muffin-tin sphere radii were chosen to be 2.2 bohr for Fe atoms and 1.8 bohr for Cl. The plane-wave cut-off energy of 12 Ry was set for the interstitial wave functions, and a $12 \times 12 \times 1$ k-mesh was sampled for integration over the Brillouin zone. To describe the electron correlation effect, several methods may be used, e.g., the local spin density approximation plus Hubbard $U$ (LSDA+$U$) method~\cite{anisimov1993}, the self interaction correction~\cite{perdew1981,shinde2021}, the hybrid functional~\cite{becke1993,andriyevsky2009}, and the GW theory~\cite{van2006}. Here the economic and practical LSDA+$U$ method was employed, with the common values of Hubbard $U$ = 4.0 eV and Hund exchange $J_{\rm H}$ = 0.9 eV for Fe $3d$ electrons. As seen below, our LSDA+SOC+$U$ calculations well reproduce the experimental band gap, which also justifies the choice of the $U$ value. The SOC is included for both Fe 3$d$ and Cl 3$p$ orbitals by the second-variational method with scalar relativistic wave functions. We critically examine the orbital states and hence determine the correct electronic ground state by calculating the crystal field level splittings and by comparing orbital multiplets using total energy calculations. For this purpose, we perform DFT calculations using spin-restricted LDA, spin-polarized LSDA, LSAD+$U$, and LSDA+SOC+$U$, as detailed below. Moreover, we perform Monte Carlo simulations on a $8 \times 8 \times 1$ spin matrix to estimate the $T_{\rm C}$ of FeCl$_2$ monolayer, using the Metropolis method~\cite{metropolis1949} and the obtained exchange parameter and magnetic anisotropy from the LSDA+SOC+$U$ calculations.

\section{RESULTS AND DISCUSSION}
\begin{figure}[t]
\centering
	\includegraphics[width=8cm]{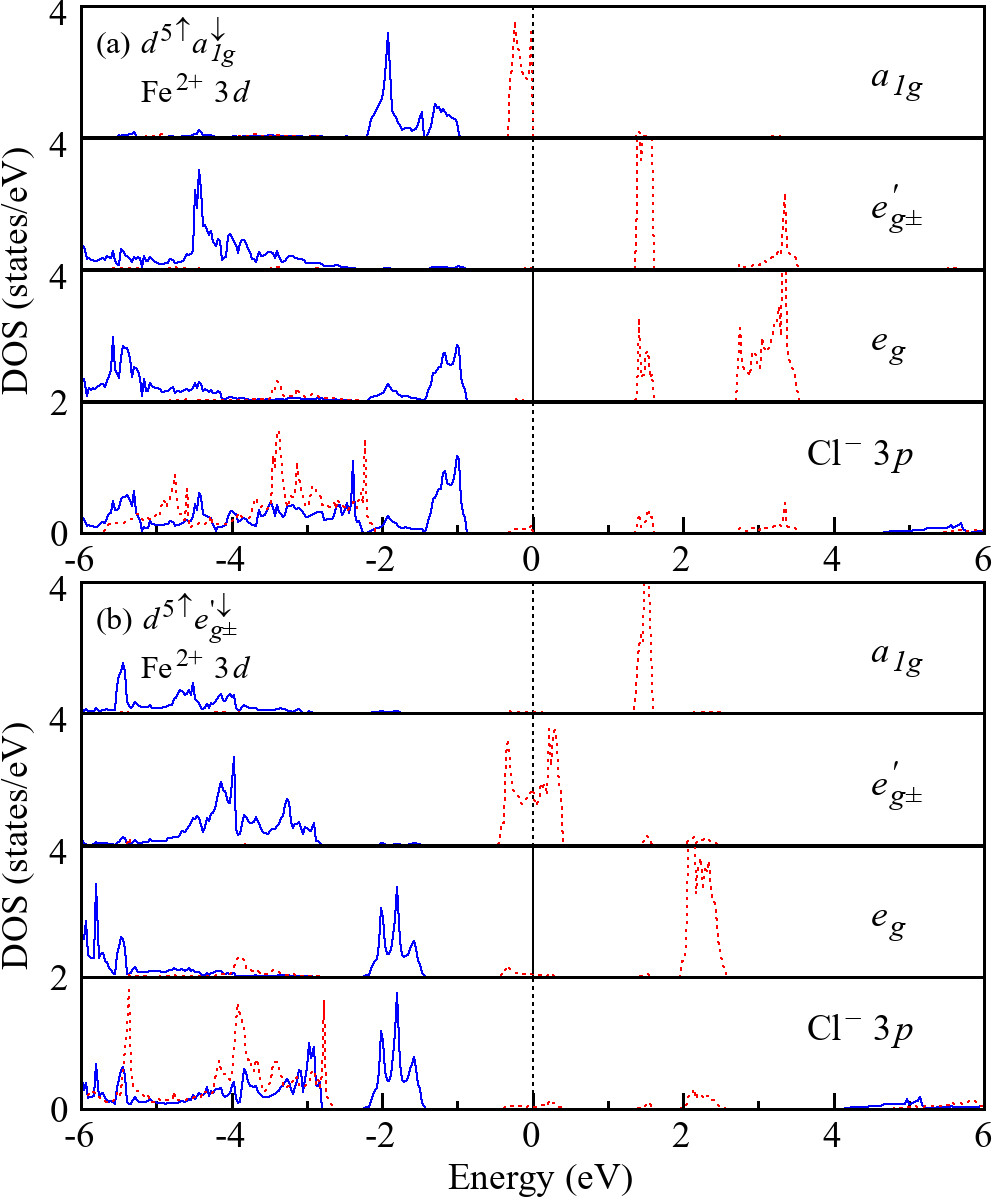}
	\caption {The DOS results of LSDA+$U$ for (a) the semiconducting $d$$^{5\uparrow}$$a$$^{\downarrow}_{1g}$ state and (b) the half-metallic $d$$^{5\uparrow}$$e'$$^{\downarrow}_{g}$ state. The blue (red) curves refer to the majority (minority) spin. The Fermi level is set at the zero energy.}
\label{fig3}
\end{figure}

We first analyze the crystal field levels of the Fe$^{2+}$ ion in FeCl$_2$ monolayer, which is crucial for understanding of the electronic ground state. As seen in Fig. 1, FeCl$_2$ monolayer adopts the 1-$T$ structure with P$\overline{3}$m1 space group~\cite{zhou2020}. Each Fe$^{2+}$ ion is surrounded by six Cl$^-$ ions, thus forming an FeCl$_6$ octahedron, and the edge-sharing octahedra comprise a triangular lattice. The local octahedral crystal field splits the Fe 3$d$ orbitals into the lower-lying $t$$_{2g}$ triplet and higher $e$$_g$ doublet. The $t$$_{2g}$ triplet further splits into the $e'_g$ doublet and the $a$$_{1g}$ singlet in the global trigonal crystal field. The global coordinate system was used in the following calculations, with the $z$ axis along the $\left[ 111 \right]$ direction of the local FeCl$_{6}$ octahedron and $y$ along the $\left[ 1\overline{1}0 \right]$ direction~\cite{wu2005,ou2014}.

To see the crystal field effect, we first perform the spin-restricted LDA calculation. Our results show that the $t$$_{2g}$-$e$$_g$ octahedral crystal field splitting is about 1 eV, see Fig. 2(a). Among the $t_{2g}$, the $a$$_{1g}$ singlet and $e'_g$ doublet are almost degenerate at a first look. In reality, by a close look the $e'_g$ is found to be slightly lower than the $a_{1g}$ by 30 meV, by calculating the center of gravity of their respective partial density of states (DOS). This result is crucial for the following discussion of the correct electronic ground state. The Cl 3$p$ state mainly lies in the range of 3-6 eV below the Fermi level, and its strong $pd\sigma$ hybridization with Fe $3d$-$e_g$ and weak $pd\pi$ hybridization with Fe $3d$-$t_{2g}$ are both evident.

Apparently, the Fe $3d$ orbitals form narrow bands in the triangular lattice, with the bandwidth less than 1 eV. Therefore, these localized $3d$ orbitals would be strongly spin-polarized by the local Hund exchange. Through the spin-polarized LSDA calculation, we find that indeed the Fe$^{2+}$ ion is in the high-spin $S$ = 2 state, see Fig. 2(b). The up-spin $3d$ orbitals are fully occupied, and the down-spin $t_{2g}$ ($a_{1g}$ and $e'_g$) is 1/3 partially occupied, seemingly giving a HM solution. Note that this HM solution is fictitious, as the strong correlation effect of the narrow bands is absent in the LSDA calculations and in reality FeCl$_2$ monolayer has a semiconducting gap of about 1.2 eV~\cite{cai2020}.

\begin{table}[t]
\small
  \caption{Relative total energies $\Delta E$ (meV/fu) for FeCl$_{2}$ monolayer in different states and under different strains by LSDA+$U$ and LSDA+SOC+$U$, and the local spin and orbital moments ($\mu_{\rm B}$) for the Fe$^{2+}$ ion. The FM state is considered in most calculations except for those marked with stripe AF, and $\perp$ ($\parallel$) represents the out-of-plane (in-plane) magnetization.}
  \label{tb1}
\setlength{\tabcolsep}{2mm}{
\begin{tabular*}{0.48\textwidth}{@{\extracolsep{\fill}}llrrrr}
\hline\hline
  & States   & $\Delta E$ & $M_{spin}$ & $M_{orb}$ \\ \hline
LSDA+$U$   &$d$$^{5\uparrow}$$a$$^{\downarrow}_{1g}$   & 0.00      & 3.55      &  $-$ \\
&$d$$^{5\uparrow}$$e'$$^{\downarrow}_{g}$   & 161.66      & 3.58      &  $-$ \\\hline
LSDA+SOC+$U$ &$d$$^{5\uparrow}$$L$$^{\downarrow}_{z+}$, $\perp$   & 0.0  & 3.56   & 0.59 \\
&$d$$^{5\uparrow}$$L$$^{\downarrow}_{z+}$, $\perp$ (AF)   & 2.83      & ${\pm3.54}$      &  ${\pm0.62}$ \\
&$d$$^{5\uparrow}$$L$$^{\downarrow}_{z+}$, $\parallel$   & 13.43      & 3.56            & 0.20   \\
& $d$$^{5\uparrow}$$a$$^{\downarrow}_{1g}$, $\parallel$   & 84.83      & 3.54     & 0.23   \\
& $d$$^{5\uparrow}$$a$$^{\downarrow}_{1g}$, $\perp$   & 91.48      & 3.54     & 0.01
\\\hline
-5$\%$   &$d$$^{5\uparrow}$$L$$^\downarrow_{z+}$, $\perp$   & 0.00      & 3.53      & 0.71 \\
&$d$$^{5\uparrow}$$L$$^\downarrow_{z+}$, $\perp$ (AF)   & 17.13      & ${\pm3.51}$      & ${\pm0.76}$ \\
&$d$$^{5\uparrow}$$L$$^\downarrow_{z+}$, $\parallel$   & 18.12      & 3.53      & 0.20 \\
&$d$$^{5\uparrow}$$a$$^{\downarrow}_{1g}$, $\parallel$   & 96.97      & 3.53      & 0.20 \\

-3$\%$   &$d$$^{5\uparrow}$$L$$^\downarrow_{z+}$, $\perp$   & 0.00      & 3.54      & 0.67 \\
&$d$$^{5\uparrow}$$L$$^\downarrow_{z+}$, $\perp$ (AF)  & 10.39      & ${\pm3.52}$      & ${\pm0.71}$  \\
&$d$$^{5\uparrow}$$L$$^\downarrow_{z+}$, $\parallel$   & 16.08      & 3.54      & 0.24 \\
&$d$$^{5\uparrow}$$a$$^{\downarrow}_{1g}$, $\parallel$   & 92.09      & 3.53      & 0.23 \\

3$\%$   &$d$$^{5\uparrow}$$L$$^\downarrow_{z+}$, $\perp$   & 2.41      & 3.57      & 0.51 \\
&$d$$^{5\uparrow}$$L$$^\downarrow_{z+}$, $\perp$ (AF)   & 0.00      & ${\pm3.56}$      & ${\pm0.54}$ \\
&$d$$^{5\uparrow}$$L$$^\downarrow_{z+}$, $\parallel$ (AF)  & 11.94      & ${\pm3.55}$      & ${\pm0.24}$ \\
&$d$$^{5\uparrow}$$a$$^{\downarrow}_{1g}$, $\parallel$ (AF)  & 68.61      & ${\pm3.52}$      & ${\pm0.32}$ \\

5$\%$   &$d$$^{5\uparrow}$$L$$^\downarrow_{z+}$, $\perp$   & 5.00      & 3.58      & 0.46 \\
&$d$$^{5\uparrow}$$L$$^\downarrow_{z+}$, $\perp$ (AF)  & 0.00      & ${\pm3.57}$      & ${\pm0.47}$  \\
&$d$$^{5\uparrow}$$L$$^\downarrow_{z+}$, $\parallel$ (AF)  & 9.11      & ${\pm3.56}$  & ${\pm0.34}$ \\
&$d$$^{5\uparrow}$$a$$^{\downarrow}_{1g}$, $\parallel$ (AF)  & 52.05      & ${\pm3.54}$      & ${\pm0.46}$
\\\hline\hline
 \end{tabular*}}
\end{table}

Now we carry out LSDA+$U$ calculations to elucidate the electron correlation effect. As seen in Fig. 3, we obtain two contrasting high-spin solutions, one with the down-spin $a_{1g}$ occupation, and the other with the down-spin $e'_g$ half filling. Apparently, the former solution has a semiconducting gap of 1.42 eV (well comparable with the experimental one of about 1.2 eV~\cite{cai2020}), while the latter one is again half metallic. Note that although the $e'_g$ doublet is lower than the $a_{1g}$ singlet in the crystal field level diagram as discussed above, the down-spin $e'_g$ doublet has to be half filled for the high-spin Fe$^{2+}$ ion due to the present symmetry constriction. Thus, this solution has to be computationally half metallic, even with strong correlation effect which is sufficient to open a band gap. Owing to the half filling of the narrow $e'_g$ band which strides over the Fermi level, this half-metallic solution is less stable, by about 162 meV/fu in our LSDA+$U$ calculations (see Table 1), than the semiconducting solution with the down-spin $a_{1g}$ occupation. The present LSDA+$U$ results show that the experimental band gap is determined by the Hubbard $U$. Then one may assume that the semiconducting solution with the down-spin $a_{1g}$ occupation is the correct one. However, this solution is an orbital singlet, and in principle it has no orbital moment but just a spin moment (3.55 $\mu_{\rm B}$/Fe, see Table 1) for the high-spin $S$ = 2 state. Even considering the SOC mixing between the nearly degenerate $a_{1g}$ and $e'_g$ (see below), a small in-plane orbital moment is expected for this solution. Then this semiconducting solution with the down-spin $a_{1g}$ occupation cannot account for the experimental perpendicular magnetic moment of 4.3 $\mu_{\rm B}$~\cite{wilkinson1959,birgeneau1972,jacobs1967}. In this sense, this semiconducting $d^{5\uparrow}a^{\downarrow}_{1g}$ solution is not the correct ground state.
\begin{figure}[t]
\centering
	\includegraphics[width=8.5cm]{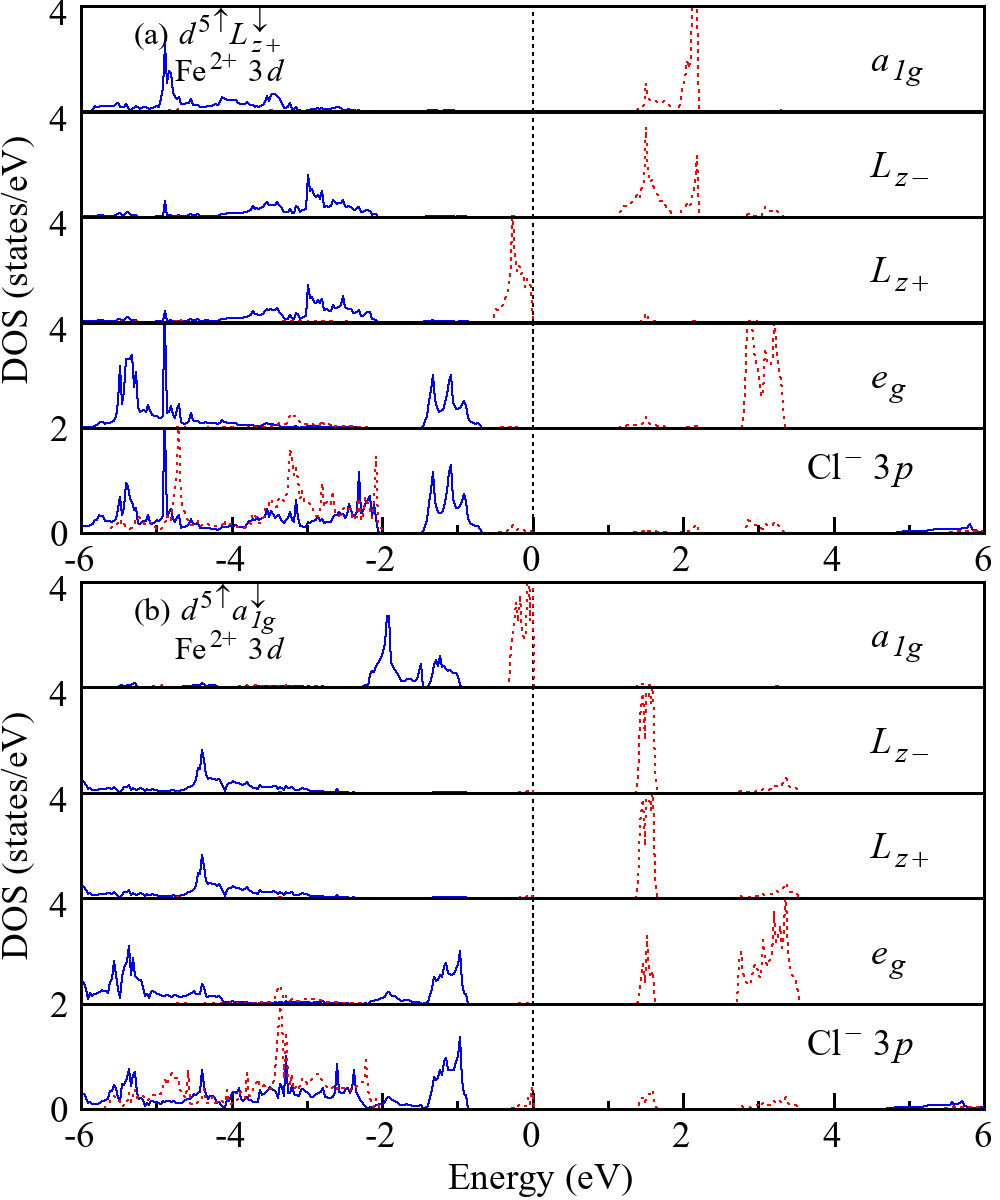}
	\caption {The DOS results of LSDA+SOC+$U$ for (a) the $d$$^{5\uparrow}$$l$$^{\downarrow}_{z+}$ ground state and (b) the $d$$^{5\uparrow}$$a$$^{\downarrow}_{1g}$ metastable state. The blue (red) curves stand for the majority (minority) spins. The Fermi level is set at the zero energy.}
\label{fig4}
\end{figure}

The Hubbard $U$ determines the band gap of FeCl$_2$ monolayer and produces a strong orbital polarization for the down-spin $t_{2g}$ ($a_{1g}$ and $e'_g$) states. In order to find the correct electronic ground state, one needs to delicately handle the orbital degrees of freedom. When the SOC is included, the $e'_g$ doublet splits into $l_{z+}$ and $l_{z-}$ states, with the respective orbital moments of +1 $\mu_{\rm B}$ and --1 $\mu_{\rm B}$ along the $z$ axis (i.e., the crystallographic $c$ axis). Then the (near) degeneracy of the $t_{2g}$ states is completely lifted upon the SOC effect, and each of them can now be subject to an orbital polarization by the Hubbard $U$. We now perform the LSDA+SOC+$U$ calculations and carefully handle the orbital multiplets, see Table 1. In particular, we now find the $d^{5\uparrow}l^{\downarrow}_{z+}$ ground state, and it is semiconducting with a band gap of 1.17 eV (very close to the experimental one of about 1.2 eV~\cite{cai2020}), see Fig. 4(a). By a simple comparison between this semiconducting solution and the LSDA+$U$ HM solution [Fig. 3(b)], one might infer that it is the SOC which opens the band gap. However, this is a wrong statement. How can the SOC (a few tens of meV in strength) open the gap more than 1 eV? Actually, here the SOC offers new orbital degrees of freedom, and the Hubbard $U$ determines the band gap. The semiconducting $d^{5\uparrow}a^{\downarrow}_{1g}$ solution remains almost unchanged by a comparison between Figs. 3(a) and 4(b), and this solution is less stable than the $d^{5\uparrow}l^{\downarrow}_{z+}$ ground state by about 85 meV/fu, see Table 1.
\begin{figure}[t]
\centering
	\includegraphics[width=9cm]{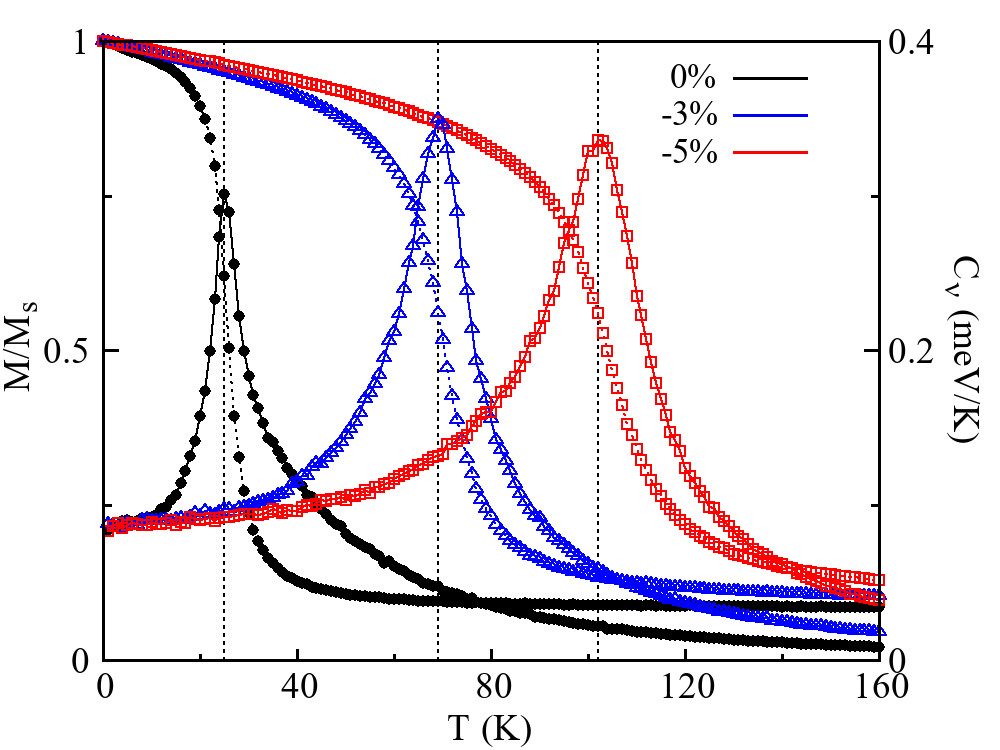}
	\caption {Monte Carlo simulations of the magnetization and the magnetic specific heat for FeCl$_{2}$ monolayer under different strains.}
\label{fig5}
\end{figure}

The $d^{5\uparrow}l^{\downarrow}_{z+}$ ground state has the orbital moment of 0.59 $\mu_{\rm B}$ along the $z$ axis, in addition to the spin moment of 3.56 $\mu_{\rm B}$. It has the easy perpendicular magnetization and is more stable than the planar magnetization by 13.43 meV/Fe, see Table 1. Therefore, the total magnetic moment of 4.15 $\mu_{\rm B}$/Fe along the $z$ axis well accounts for the experimental perpendicular moment of 4.3 $\mu_{\rm B}$~\cite{wilkinson1959,birgeneau1972,jacobs1967}. Moreover, the $d^{5\uparrow}l^{\downarrow}_{z+}$ ground state prefers a FM coupling in FeCl$_2$ monolayer, and it is more stable than the stripe AF state by 2.83 meV/Fe. Using the magnetic exchange expressions $-JS^2$ for each FM pair with $S$ = 2, $-3JS^2$ per fu for the FM state, and $JS^2$ per fu for the stripe AF state, we derive the FM exchange parameter $J$ = 2.83/4$S^2$$\approx$ 0.18 meV. In addition,
for the metastable $d^{5\uparrow}a^{\downarrow}_{1g}$ state, the SOC mixes the nearly degenerate $a_{1g}$ and $e'_g$ states and produces a planar orbital moment of 0.23 $\mu_{\rm B}$. Note that this easy planar magnetization would fail to explain the experimental perpendicular magnetization. Therefore, all the above LSDA+SOC+$U$ results lead us to a conclusion that $d^{5\uparrow}l^{\downarrow}_{z+}$ (but not $d^{5\uparrow}a^{\downarrow}_{1g}$) is the correct ground state and it consistently explains the FM semiconducting behavior of FeCl$_2$ monolayer with the perpendicular magnetization.

In order to estimate the $T_{\rm C}$ of FeCl$_2$ monolayer, we assume the spin Hamiltonian and carry out Monte Carlo simulations
\begin{equation*}
\begin{aligned}
H=-\frac{J}{2} \sum_{\langle i j\rangle} \overrightarrow{S_{i}}\cdot \overrightarrow{S_{j}}-D \sum_{i}(S^z_i)^{2},
\end{aligned}
\end{equation*}
where the first term represents the isotropic Heisenberg exchange with $J=0.18$ meV, and the $D$ parameter in the second term stands for the magnetic anisotropy of the single ion type. Our LSDA+SOC+$U$ results listed in Table 1 suggest the easy $z$-axis magnetization with the anisotropy energy of 13.43 meV/Fe and $S_z = 2$, and thus $D=3.36$ meV can be derived. Using these $J$ and $D$ parameters, our Monte Carlo simulations give $T_{\rm C}$ = 25 K for the pristine FeCl$_2$ monolayer, see Fig. 5.
\begin{figure}[t]
\centering
	\includegraphics[width=9cm]{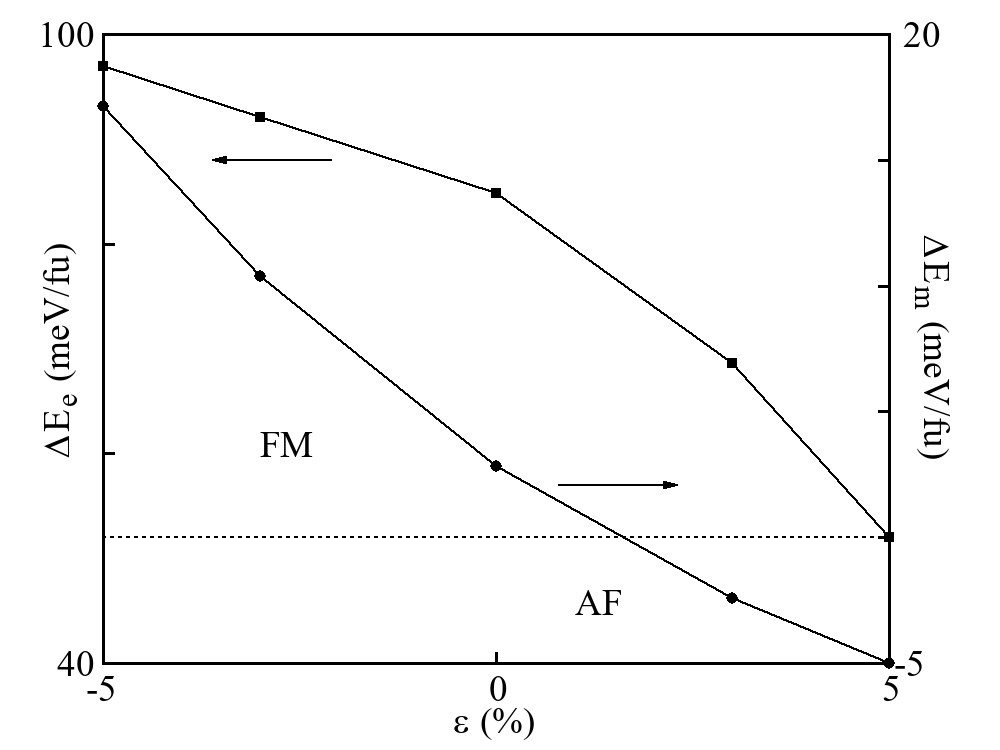}
	\caption {$\Delta E_{\rm e}$ (meV/fu): the relative stability of the $d$$^{5\uparrow}$$l$$^\downarrow_{z+}$ ground state against the $d$$^{5\uparrow}$$a$$^{\downarrow}_{1g}$ state under different strains. $\Delta E_{\rm m}$ (meV/fu): the $d$$^{5\uparrow}$$l$$^\downarrow_{z+}$ FM state against the AF state.}
\label{fig6}
\end{figure}

As a lattice strain is an effective way to tune the electronic state and magnetism of 2D materials, here we also study a biaxial strain effect on the FM of FeCl$_2$ monolayer. A compressive strain would raise the crystal field level of the $a$$_{1g}$ singlet, and thus further stabilizes the $d$$^{5\uparrow}$$l$$^\downarrow_{z+}$ ground state. As shown in Table 1 and Fig. 6, the 3\% or 5\% compressive strain gives a larger orbital moment, a stronger perpendicular anisotropy, and a stronger FM coupling, according to the LSDA+SOC+$U$ calculations. Therefore, the $T_{\rm C}$ is naturally expected to be enhanced, and this is indeed confirmed by our Monte Carlo simulations (see Fig. 5): $T_{\rm C}$ is 69 K for the --3\% strain and 102 K for --5\% strain. In addition, we study a tensile strain, which would gradually destabilize the $d$$^{5\uparrow}$$l$$^\downarrow_{z+}$ ground state. Our LSDA+SOC+$U$ calculations show that the orbital moment gets smaller, the perpendicular magnetic anisotropy shrinks, and strikingly, the intralayer magnetic coupling changes its sign: the stripe AF state gets more stable than the FM state upon 3\% and 5\% strain. All these results show that a compressive strain would significantly enhance the $T_{\rm C}$ of FeCl$_2$ monolayer, but that a tensile strain could trigger an interesting FM-AF transition. Therefore, FeCl$_2$ monolayer could be an appealing 2D magnetic semiconductor potentially suitable for spintronic applications.

\section{CONCLUSIONS}
In summary, using a set of delicate DFT calculations including the SOC and Hubbard $U$, aided with the crystal field level analyses, we achieve the correct $d^{5\uparrow}l^{\downarrow}_{z+}$ ground state for FeCl$_2$ monolayer. This ground state well explains, in a consistent way, the experimental FM semiconducting behavior with the perpendicular magnetization, and our Monte Carlo simulation gives $T_{\rm C}$ = 25 K for the pristine FeCl$_2$ monolayer. Moreover, we find that upon the compressive strain, the $d^{5\uparrow}l^{\downarrow}_{z+}$ ground state gets more stable, and the enhanced FM coupling and the perpendicular magnetic anisotropy raise the $T_{\rm C}$ a lot, up to 69 K (102 K) for --3\% (--5\%) strain. We also predict an interesting FM-AF transition for FeCl$_2$ monolayer under a tensile strain. All these results suggest that FeCl$_2$ monolayer is an appealing 2D magnetic semiconductor.

\section{ACKNOWLEDGEMENTS}
This work was supported by National Natural Science Foundation of China (Grants No. 12104307 and No. 12174062).

\bibliography{rsc}
\end{document}